\documentclass[11pt,twoside]{article}

\usepackage{asp2006}
\usepackage{epsf}
\usepackage{psfig}
\usepackage{lscape}

\newcommand{\p}[2]{\frac{\partial #1}{\partial #2}}
\newcommand{\od}[2]{\frac{d #1}{d #2}}
\newcommand{\h}[1]{{\cal#1}}

\begin{document}
\title{Calculation of Solar P-mode Oscillation Frequency Splittings 
Based on a Two-dimensional Solar Model}   
\author{Linghuai Li, Sarbani Basu, Sabatino Sofia, and Pierre Demarque}   
\affil{Department of Astronomy, Yale University, P.O. Box 208101, New Haven, CT 06520-8101}    

\begin{abstract} 
We compute the p-mode oscillation frequencies and frequency splittings that arise in a two-dimensional model of the Sun that contains toroidal magnetic fields in its interior.
\end{abstract}


\section{Introduction and Basic Equations}   

Rotation and magnetic fields in the solar interior result in asphericity. This asphericity 
results in solar p-mode oscillation frequency splittings. 
From our 2D code we are able to obtain 2D solar models. In order to do helioseismic
tests of the 2D effects, we need to calculate the p-mode frequency from the 2D solar
models. The question is how to do so.

No doubt, the starting points are the MHD equations. Nevertheless, it would be convenient to 
use the same format as in \S19 of Unno et al. (1989). We reformulate them by defining 
the total pressure by adding the isotropic magnetic pressure component to the gas pressure, 
as we did in our 1D and 2D solar variability model calculations (Lydon and Sofia 1995; 
Li et al. 2001, 2002, 2003, 2006). The zero-order approximation of the MHD equations is equivalent
to our stellar structure equations, which give rise to the equilibrium model of the Sun. We can
obtain 1D or 2D solar model by using our 1D and 2D codes.

The first-order approximation of the MHD equations yields the pulsation equations. 
The following adiabatic oscillation equations hold well for any equilibrium state when the 
magnetic fields are present, regardless its symmetry:
\begin{eqnarray}
&& \nabla\cdot{\bf{\xi}} = -\frac{1}{\Gamma_1 P}(p'+\xi\cdot\nabla P), \nonumber \\
&& \sigma^2\xi+\nabla\phi'+\frac{1}{\rho}\nabla p' +\frac{\rho'}{\rho}\nabla\phi =\h{H}', 
\label{eq:osci} \\
&& \frac{1}{r^2}\p{}{r}(r^2\p{\phi'}{r})+\nabla^2_\perp\phi' = 4\pi G\rho', \nonumber
\end{eqnarray}
where
\begin{eqnarray*}
 \h{H}' &=&\frac{1}{4\pi\rho}\left\{(\bf{B}\cdot\nabla)\bf{B}'+({B}'\cdot\nabla)\bf{B}\right\},\\
  \bf{B}'&=& \nabla\times(\xi\times\bf{B}).
\end{eqnarray*}
The symbols without primes represent the equilibrium state, which can either be 
spherically-symmetric, as for the 1D stellar models, or aspherical, as for the 2D stellar models
that are to be extended in this paper. The symbols with primes are the Euler perturbations. 
It is well-known that $\xi$ represents the Lagrangian displacement vector.

The adiabatic oscillation equations contain four equilibrium state variables. They are pressure (P),
density ($\rho$), gravitational potential ($\Phi$) and the adiabatic index ($\Gamma_1$). All of them
are functions of radial and colatitudinal coordinates ($r,\theta$) in the azimuthal symmetry. 
We want to decompose them into two components in order to extend the oscillation equations into
two dimensions like this:
\begin{equation}
  P(r,\theta)=P(r) + p_1(r,\theta). \label{eq:decomp}
\end{equation}
Here we assume that $P(r)$ is the mean value of pressure over $\theta$.
As a result, $p_1(r,\theta)$ is the pressure asphericity. Similar for the others.
This decomposition will automatically reduce to its 1D counterpart when the equilibrium 
structure is spherically-symmetric.

Substituting Eq.~(\ref{eq:decomp}) into Eqs.~(\ref{eq:osci}) we obtain two-dimensional 
adiabatic oscillation equations:
\begin{eqnarray}
 && \frac{1}{r^2}\p{}{r}(r^2\xi_r)+\frac{1}{\Gamma_1}\od{\ln P}{r}\xi_r
    +\left(\frac{\rho}{\Gamma_1 P}
    +\frac{\nabla^2_\perp}{\sigma^2}\right)\frac{p'}{\rho}
    +\frac{1}{\sigma^2}\nabla_\perp^2\Phi' \nonumber \\
 &&  = - \h{A}  - \frac{1}{\sigma^2}\nabla_\perp\cdot(\h{B}_\perp-\h{H}'_\perp), \nonumber \\
&& \frac{1}{\rho}\left(\p{}{r}+\frac{\rho g}{\Gamma_1 P}\right)p'
    -(\sigma^2+gA)\xi_r+\p{\Phi'}{r} = -g\h{C}-(\h{B}_r-\h{H}'_r) \label{eq:general} \\
&& \left(\frac{1}{r^2}\p{}{r}r^2\p{}{r}+\nabla_\perp^2\right)\Phi'-4\pi 
    G\rho\left(\frac{p'}{\Gamma_1 P}-A\xi_r\right) = 4\pi G\rho \h{C}. \nonumber
\end{eqnarray}
The left hand sides are the same as their 1d counterparts.
The right hand sides represent the 2d effects, in which
\begin{eqnarray*}
 \h{A} &=& \frac{\xi\cdot\nabla p_1}{\Gamma_1 P}-\frac{1}{\Gamma_1}\left(\frac{p'}{P}+\xi_r\od{\ln P}{r}\right) \left(\frac{\Gamma_{1\,1}}{\Gamma_1}+\frac{p_1}{P}\right), \\
 \h{C} &=& \frac{\xi\cdot\nabla p_1}{\Gamma_1 P}-\frac{1}{\Gamma_1}\left(\frac{p'}{P}+\xi_r\od{\ln P}{r}\right) \left(\frac{\Gamma_{1\,1}}{\Gamma_1}+\frac{p_1}{P}-\frac{\rho_1}{\rho}\right), \\
 \h{B}_r &=& -\frac{\rho_1}{\rho^2}\p{p'}{r} 
   +\left(\frac{\rho_1g}{\rho}-\p{\Phi_1}{r}\right)\left(A\xi_r
   -\frac{p'}{\Gamma_1P}\right),\\
 \h{B}_\perp &=& -\frac{\rho_1}{\rho^2}\nabla_\perp p'  
   -(A\xi_r-\frac{p'}{\Gamma_1 P})\nabla_\perp\Phi_1.
\end{eqnarray*}

In order to solve these equations we treat the two-dimensional 
correction terms like this: (1) to expand the oscillation variables in 
the 2D corrections in terms of spherical harmonics; (2) to remove the 
angular dependence of the 2D corrections this way:
(a) Multiplying them by the complex conjugate of the spherical harmonics;
(b) Integrating them over the solid angle;
(c) Summing them over all degree l;
(d) Summing them over all azimuthal order m.

\section{Two-dimensional solar models and frequency splittings}

The torus field is defined as follows: $B_\phi=B_0\sqrt{\rho/\rho_0}$ in the torus tube
$[(R-2a)\sin\theta_0-r\sin\theta]^2+[r\cos\theta-(R-2a)\cos\theta_0]^2=a^2$, but $B_\phi=0$ 
outside the tube, where $a$ is the tube radius, $c=(R-2a)\sin\theta_0$ is the distance of the tube
center to the pole axis, $\theta_0$ is the colatitude of the tube center, $R$ is the solar radius,
$\rho_0$ is the density at the solar surface, $B_0$ is the magnetic strength parameter.

\begin{figure}[!ht]
\plotone{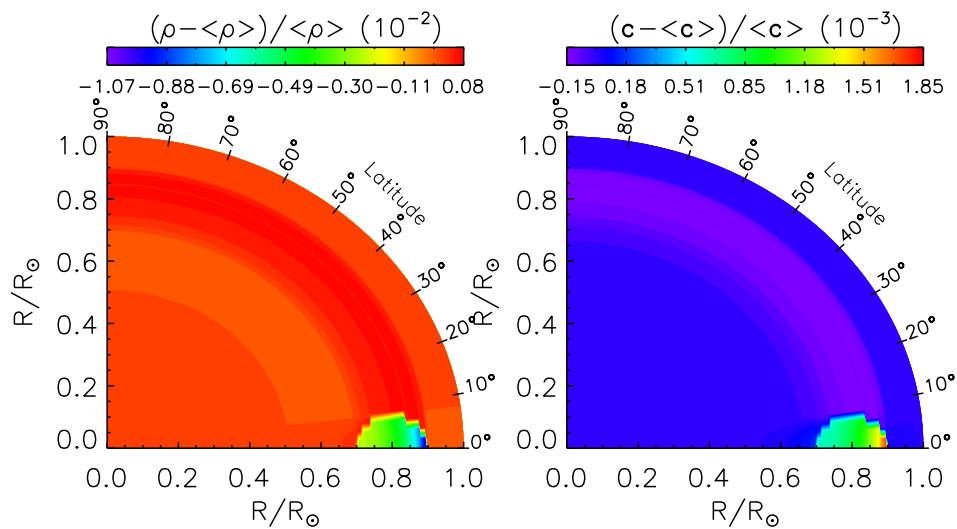}
\caption{Dependence of density (left) and sound speed (right) asphericities 
on radius and colatitude in Model 1}\label{fig:torus1}
\end{figure}

We want to investigate three 2D solar models: Model 1.  $a=70$ Mm, $\theta_0=90^\circ$, 
$B_0=2400$, $B_{max}=2.19$ MG, resolution $2576\times44$; Model 2. The same as Model 1 except for
resolution $2576\times90$; Model 3. The same as Model 2 except for resolution $2576\times180$.
The 1D standard solar model (SSM) is also needed to compute the central frequency 
differences between 2D and 1D models. Fig.~\ref{fig:torus1} shows the density (left) 
and sound speed (right) asphericities in Model 1.

\begin{figure}[!ht]
\plottwo{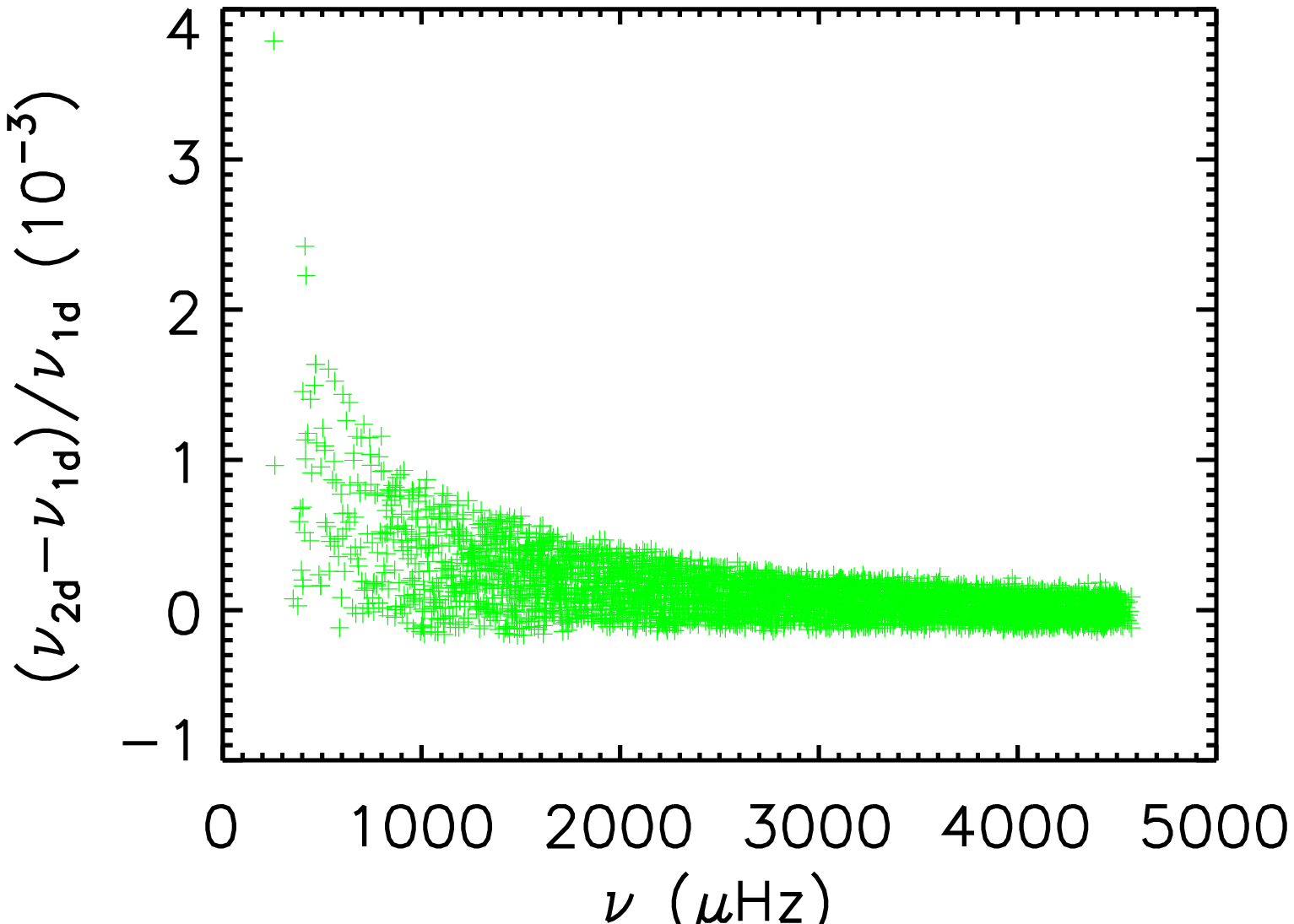}{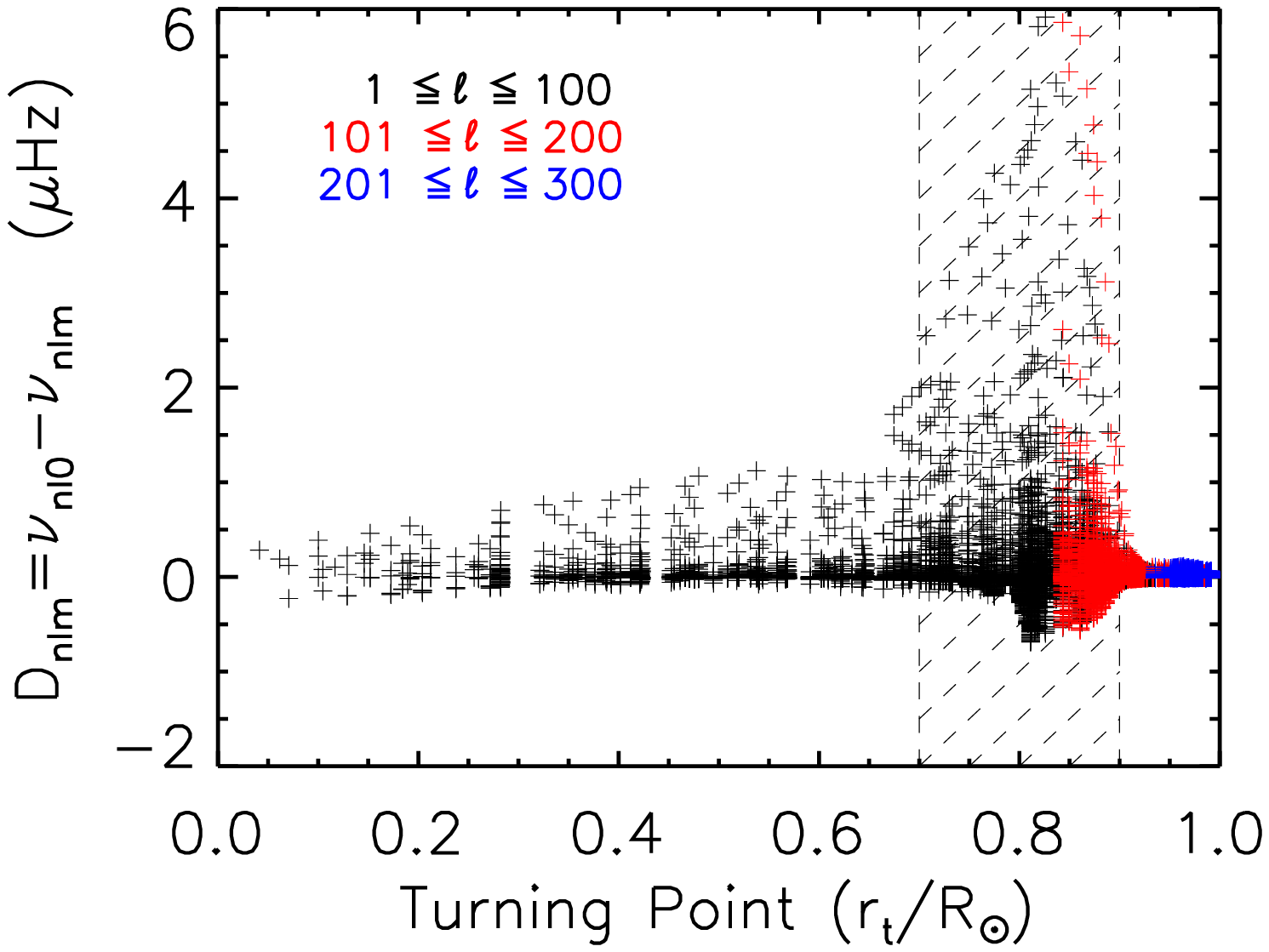}
\caption{Relative central frequency difference (left) of 2D model 1 with 
respect to SSM. Dependence of frequency splittings (right) on the inner turning 
points for Model 1}\label{fig:difsplit}
\end{figure}

The central frequency is defined as $\nu_{nl}=\nu_{nlm=0}$, and the frequency splitting is 
defined as $D_{nlm} = \nu_{nlm=0}-\nu_{nlm}$.
It is interesting to investigate how they change with the inner turning points. 
Fig.~\ref{fig:difsplit} shows the relative central frequency difference (left) of 2D model 1 with 
respect to SSM and the dependence of Frequency splittings on the inner 
turning points for Model 1 (right). Obviously, the splittings for 
whose turning points locate within the tube peak. This is easy to understand:
their waves spend most time near their turning points since their group velocity 
vanishes at the turning points.

\begin{figure}[!ht]
\plottwo{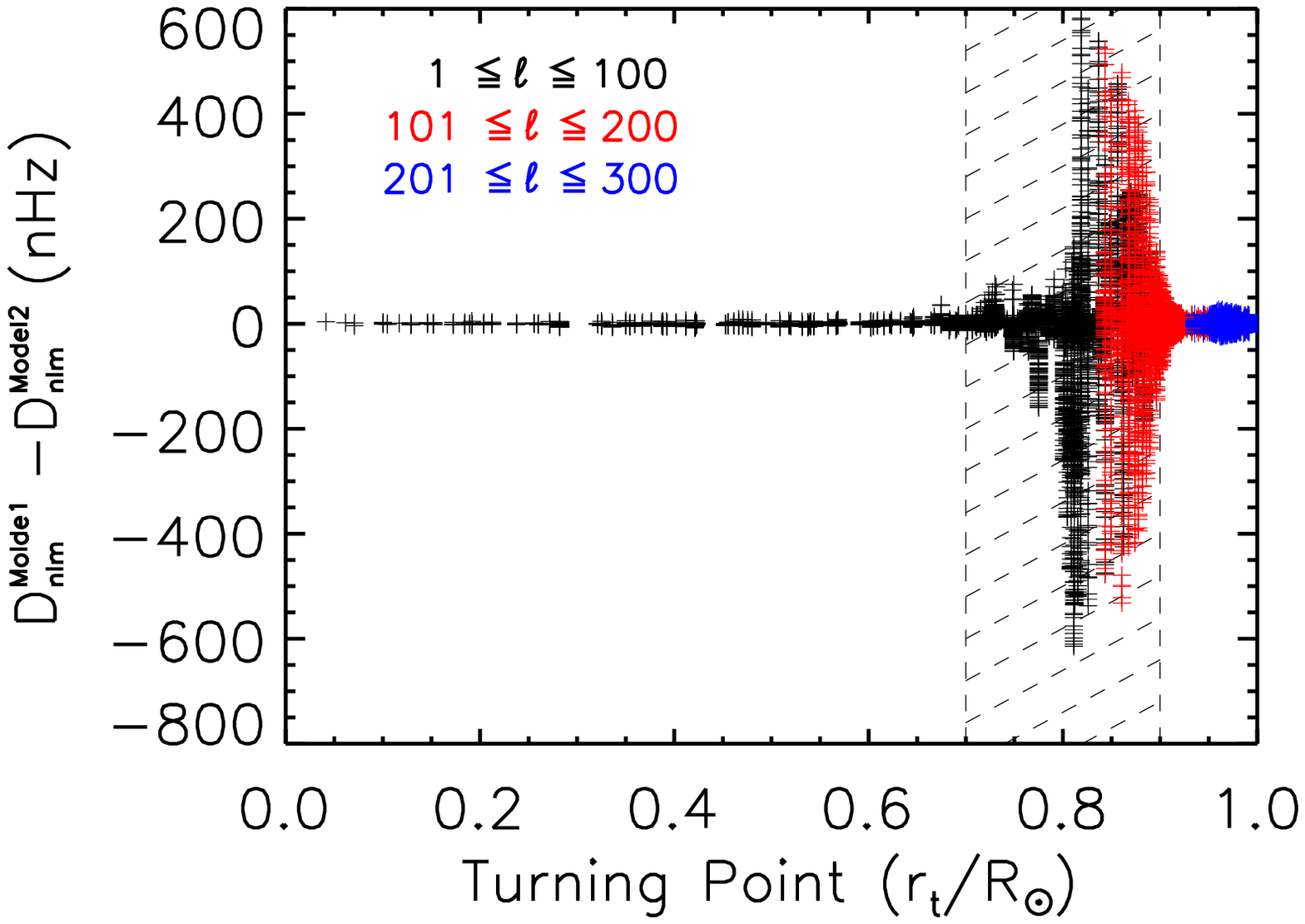}{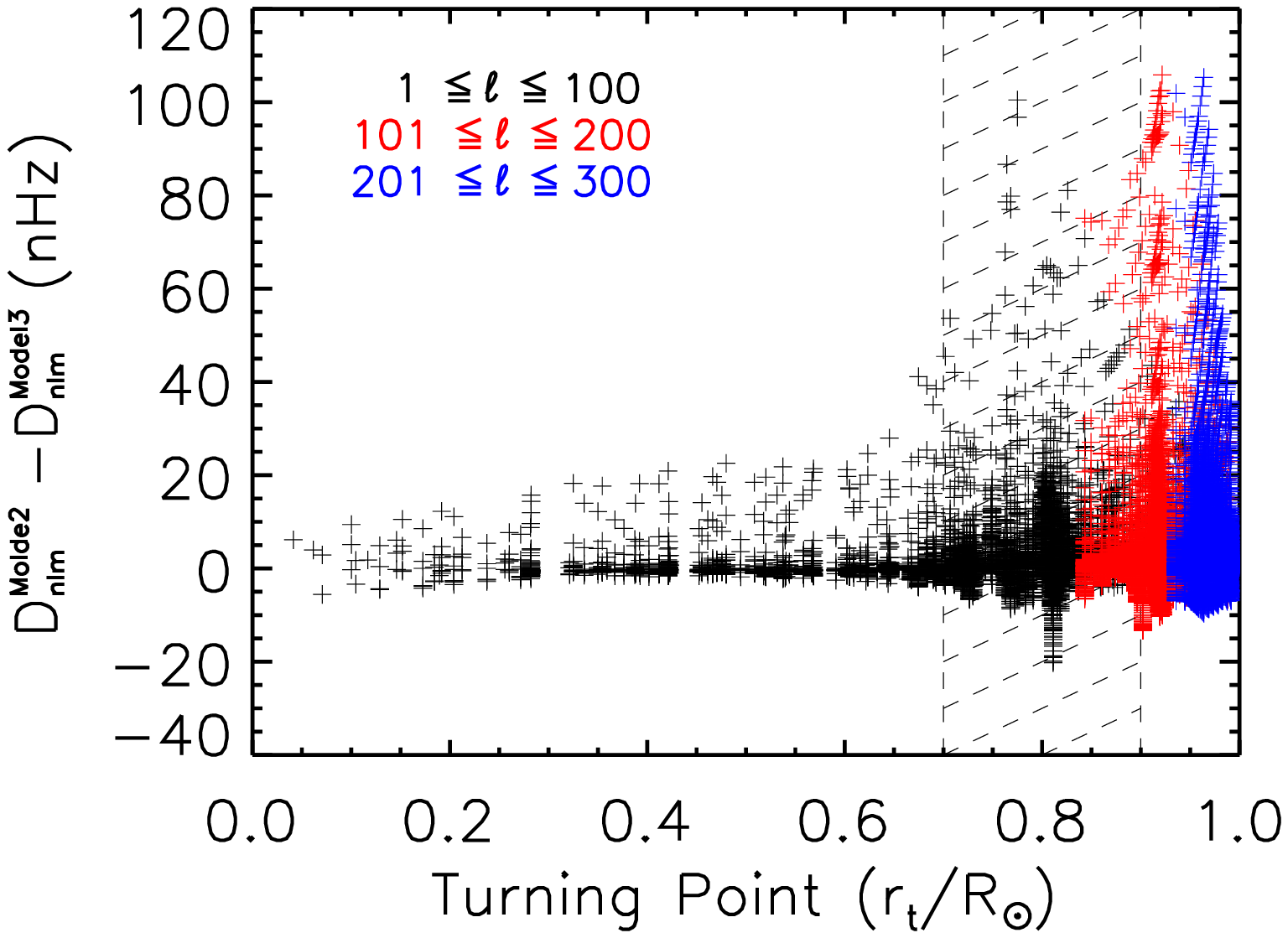}
\caption{Dependence of splitting differences between Models 2 and 3 (left) and 3 and 4 (right) 
on the inner turning points. The shadow region is the location of the torus tube.}\label{fig:split1}
\end{figure}

Models 2 and 3 are designed to check how high the angular resolution is enough to minimize 
the numerical errors in the frequency splitting calculations. Fig.~\ref{fig:split1} shows 
the splitting differences between Models 1 and 2 (left) and Models 2 and 3 (right). 
Model 3 doubles the angular resolution of Model 2, which doubles that of Model 1. The fact 
that the splitting differences between Models 2 and 3 decrease in comparison with those 
between Models 1 and 2 demonstrates that the angular resolution for Model 3 is not 
high enough yet. The main cause is that the step function gives rise to discontinuity. 

We come to the following tentative conclusion:
The frequency splittings change with the turning point this way:
(a) the biggest when it is in the torus tube,
(b) the smallest when it is above the tube,
(c) in between when it is below the tube.
We need to use a smooth field in order to obtain splittings of the required precision.

\acknowledgements We want to acknowledge the support for LLH by NSF Grant 
ATM 073770 and the Vetlesen Foundation; SS by the Vetlesen and the Brinson 
Foundations; SB by NSF grants ATM 0348837 and ATM 0737770.


\end{document}